\documentclass[aps,prx,twocolumn,superscriptaddress,longbibliography]{revtex4-1}
\usepackage{graphicx}
\usepackage{epstopdf}
\usepackage{amsmath}
\usepackage{amssymb}
\usepackage{color}
\usepackage[latin1]{inputenc}
\DeclareMathOperator{\arsinh}{arsinh}
\DeclareMathOperator{\artanh}{artanh}
\begin{document}
\renewcommand{\vec}[1]{\boldsymbol{#1}}

\title{Electro-osmotic properties of porous permeable films}

\author{Elena F. Silkina}

\affiliation{Frumkin Institute of Physical Chemistry and Electrochemistry, Russian Academy of Sciences, 31 Leninsky Prospect, 119071 Moscow, Russia}

\author{Naren Bag}
\affiliation{DWI - Leibniz Institute for Interactive Materials,  Forckenbeckstr. 50, 52056 Aachen, Germany}

\author{Olga I. Vinogradova}
\email[Corresponding author: ]{oivinograd@yahoo.com}

\affiliation{Frumkin Institute of Physical Chemistry and Electrochemistry, Russian Academy of Sciences, 31 Leninsky Prospect, 119071 Moscow, Russia}
\affiliation{DWI - Leibniz Institute for Interactive Materials,  Forckenbeckstr. 50, 52056 Aachen, Germany}
\date{\today}

\begin{abstract}
Permeable porous coatings on a flat solid support significantly impact its electrostatic and electrokinetic properties.
Existing work has focused on simplified cases, such as weakly charged and/or thick porous  films, with limited theoretical guidance.
Here, we consider the general case of coatings of an arbitrary volume charge density and obtain analytic formulas for electrostatic potential profiles,  valid for any film thickness and salt concentration.  Our analysis provides a framework for interpreting
and predicting super
properties specific to porous films,  from an enhanced ion absorption to a consequent amplification of
electro-osmotic flows due to emergence of slip velocity at interface with an outer electrolyte leading to a large zeta-potential. The latter can be tuned by varying amount of added salt, and remains finite at even high  concentrations.
The results are relevant for hydrogel coatings,  porous carbon and silica, polyelectrolyte brushes, and more.
\end{abstract}

\maketitle

\section{Introduction}

 Charged porous materials that are permeable to water and ions, such as polyelectrolyte networks, ion-exchange resins, silica gels,  porous membranes and electrodes, have found use in a large body of  applications including water desalination~\cite{porada.s:2012}, tissue engineering~\cite{stamatialis.df:2008, chung.hh:2018}, and electrochemical systems~\cite{biesheuvel.pm:2011}. Thanks to a recently discovered  extremely strong electrokinetic flow near porous surfaces~\cite{feldmann.d:2020}, new opportunities  in microfluidics and advanced colloid technologies are emerging.
Porous films on a variety of supports, are similarly capable to provide such properties as improved transport and storage capacities for ions,  that they did not have when impermeable. However, the quantitative understanding of novel equilibrium and transport properties, which could not be achieved without porosity is still challenging.


A considerable progress has been made over the last decades in understanding the equilibrium properties of porous surfaces in electrolyte solutions. Analytic solutions based on a linearized Poisson-Boltzmann theory are known~\cite{donath.e:1979,ohshima.h:1985,ohshima.h:1990a,chanda.s:2014}, but these results do not apply to highly charged coatings, where nonlinear electrostatic effects could become significant. The non-linear electrostatic problem has been  treated using numeric and semi-analytic  approaches~\cite{ohshima.h:1985,duval.jfl:2005,chen.g:2015}, and some simple analytic expressions for the static surface potential, $\Psi_s$ have  been derived for thick coatings compared to the Debye screening length, $\lambda_D$, which is a measure of the thickness of the electrostatic diffuse
layer~\cite{ohshima.h:1985,silkina.ef:2020}. Nevertheless,  approaches to calculate $\Psi_s$ analytically are sill lacking and general principles to control it have not yet been established, especially for the case of strongly  charged coatings of a thickness smaller or comparable to $\lambda_D$.

\begin{figure}
\begin{center}
\includegraphics[width=0.9\columnwidth]{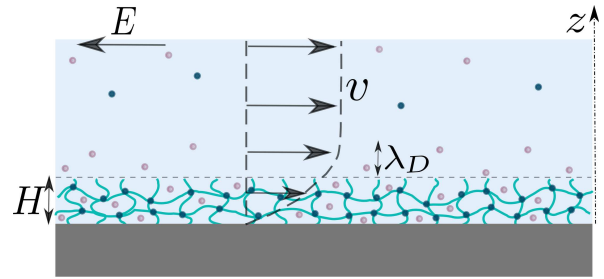}
\end{center}
\vspace{-0.4cm}
\caption{Porous film of thickness $H$ in contact with an electrolyte solution.  Anions and cations are denoted with
bright and dark circles. An outer electrostatic diffuse layer of a thickness of the order of $\lambda_D\equiv \kappa^{-1}$, is formed in the vicinity of the coating. A tangential electric field, $E$, leads to
a solvent flow of velocity $v$ (shown by right arrows).} \label{Fig:sketch}
\end{figure}

The electro-kinetic properties of porous surfaces in electrolyte solutions are relatively less understood, although there is some literature describing attempts to provide a theory of electro-osmosis near porous surfaces. It has been understood that $\Psi_s$ of porous surfaces does not define unambiguously the electro-osmotic flow properties~\cite{donath.e:1979}, and that bulk velocity is controlled, besides $\Psi_s$, by the Brinkman and inner Debye screening lengths~\cite{ohshima.h:1990a}.
These authors, however, failed to propose a physical interpretation of their results, which generally indicate that porous surfaces can amplify electro-osmotic pumping  at micron scales, where pressure-driven flows are suppressed by viscosity. These theories and subsequent attempts at improvement the description of the electroosmosis near porous surfaces~\cite{ohshima.h:1995,duval.jfl:2004,duval.jfl:2005,ohshima.h:2006} are often invoked in the interpretation of the electro-osmotic data, but their relation to electrokinetic (zeta) potential, $Z$, which is the measure of electrokinetic mobility, has remained somewhat obscure. Some authors concluded that it `loses its significance'~\cite{ohshima.h:1990a} or `is undefined and thus nonapplicable'~\cite{duval.jfl:2004}, while others reported that $Z$ typically exceeds $\Psi_s$~\cite{sobolev.vd:2017,chen.g:2015}, but did not attempt to relate \textbf{their} results to the inner flow and emerging liquid velocity at the interface. Besides, neither paper attempted to understand an upper bound on achievable zeta-potential. We, therefore, gained the impression that certain crucial aspects of the electrokinetics near the porous interface are either poorly understood or have been given so far insufficient attention.

In this paper, we describe analytically the distribution of a potential induced by a planar porous coating. Our simple analytic expressions are valid even when the volume charge density is quite large, and can be used for any salt concentration and  thickness of a porous layer. From this theory, we are interpreting enhanced absorption properties of porous films, and show that due to these mobile absorbed ions the electro-osmotic flow inside the porous film emerges, which in turns leads to   the finite slip velocity at the porous interface. The later is the reason for an enhancement of the electro-osmotic velocity in the bulk electrolyte and of zeta-potential. Finally, we obtain an upper bound on attainable zeta-potential that provides guidance for a giant amplification of electro-osmotic flows.

\section{Electro-osmotic equilibria}

The system geometry is shown in Fig.~\ref{Fig:sketch}. The permeable film of a thickness $H$ and a fixed volume charge density $\varrho$ (taken positive without loss of generality) is placed in an 1:1 electrolyte solution of bulk concentration $c_{\infty}$ and permittivity $\varepsilon$. Ions obey Boltzmann distribution, $c_{\pm }(z)=c_{\infty}\exp (\mp \psi (z))$, where $\psi (z)=e \Psi (z)/(k_{B}T)$ is the dimensionless
electrostatic potential, $e$ is the elementary positive charge, $k_{B}$ is
the Boltzmann constant, $T$ is a temperature, and the upper
(lower) sign corresponds to the cations (anions). The inverse Debye screening length of an electrolyte solution, $\kappa \equiv \lambda_D^{-1}$, is defined as usually, $\kappa^2=8 \pi \ell_B c_{\infty}$, with the Bjerrum length $\ell_B=\dfrac{e^2}{\varepsilon k_BT}$.

\subsection{Electrostatic Potentials}

To calculate the profile of a potential inside the porous  film and in the outer solution we
 employ the nonlinear Poisson-Boltzmann theory~\cite{andelman.d:1995}, so that $\psi (z)$ satisfies
\begin{equation}
\psi_{i, o}^{''} = \kappa^2 \left(\sinh\psi_{i, o} -\rho\Theta \left(H - z  \right)\right) \label{Eq:PB_io},
\end{equation}
where $^{\prime}$ denotes $d/d z$, with the index $\{i, o\}$ standing for ``in'' $(z \leq H)$ and ``out'' $(z \geq H)$, $\rho = \dfrac{\varrho}{2ec_{\infty}}$, and  the Heaviside step function $\Theta(z)$. The solution of non-linear Eq.\eqref{Eq:PB_io} with prescribed boundary conditions, in  general requires  a  numerical method. Here we obtain a closed-form analytical solution.

Integrating Eq.\eqref{Eq:PB_io} twice by applying conditions $\psi_o^{'}\rightarrow 0$ and $\psi_o \rightarrow 0 $ at $z\rightarrow \infty$ we naturally find that the $\psi_o-$profile is identical to derived for an impenetrable wall~\cite{andelman.d:1995}
\begin{equation}\label{eq:PB_out3}
\psi_{o} =  4 \artanh \left[ \gamma e^{-\kappa (z-H)}\right]
\end{equation}
where $\gamma =\tanh\dfrac{\psi_s}{4}$ and $\psi_s = \psi (H)$ is the surface potential. Using $\psi_i^{'} (0) = 0$ and $\psi_i (0)=\psi_0$ we obtain
that $\psi_0$  and $\psi_s$ are related as
\begin{equation}\label{eq:psi_s_thick}
\psi_s \equiv \psi_0 - \frac{\cosh \psi_0 - 1}{\rho},
\end{equation}
derived before for thick films only~\cite{silkina.ef:2020}.
Note that Eqs.\eqref{eq:psi_s_thick} and \eqref{eq:PB_out3}, are exact and valid for any $\kappa H$ and $\rho$.

Further insight can be gained by recalling that  $\cosh \psi$ represents a dimensionless local osmotic pressure of an electrolyte solution, $p=P/2 c_{\infty} k_B T$, which takes its largest value of $p_0 = \cosh \psi_0$ at $z = 0$. Since $p (\infty)=1$, Eq.\eqref{eq:psi_s_thick} indicates that an excess osmotic pressure at the wall grows linearly with a self-induced  potential difference across the porous film, $\psi_0 - \psi_s$. It will be clear below that this is a main parameter that ascertains most of its properties.

In the limit of a thin film, $\kappa H \ll 1$, the asymptotic approach suggested before \cite{silkina.ef:2019} can be employed.
Expanding the potential in Eq.\eqref{Eq:PB_io} about $z=0$, we obtain, to second order in $\kappa z$
\begin{equation}\label{eq:phi_in_thin}
\psi_{i}(z)\simeq \psi_0 - \dfrac{\rho}{2} \left(\kappa z\right)^2 \left[ 1 - \frac{\sinh \psi_0}{\rho} \right].
\end{equation}
Note that the value of $ \sinh \left(\psi_0 \right)/\rho \leq 1$ represents the degree of screening of the film intrinsic charge at $z = 0$.

\begin{figure}[t]
\begin{center}
\includegraphics[width=1\columnwidth]{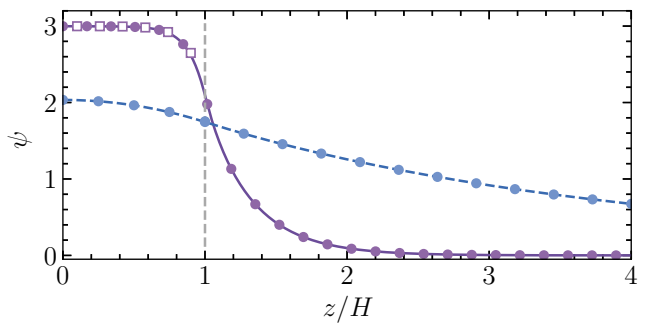}
\end{center}
\vspace{-0.4cm}
\caption{A distribution of a potential built up  by a film of $\rho=10$, calculated numerically for $\kappa H = 0.3$ (dashed curve) and $3$ (solid curve).  Filled circles correspond to calculations from Eqs.\eqref{eq:phi_in_thin1}, \eqref{eq:new}, when $z/H \leq  1$, and from Eq.\eqref{eq:PB_out3}, when $z/H \geq  1$. Open squares are obtained from Eq.\eqref{eq:phi_in_thick2}.} \label{fig:Fig2}
\end{figure}

From Eqs.\eqref{eq:phi_in_thin} and \eqref{eq:psi_s_thick} it follows that
$\psi_0$ satisfies
\begin{equation}\label{eq:rho_of_psi}
    \rho^2 - \rho \sinh \psi_0 -\frac{2 \left(\cosh \psi_0 - 1 \right)}{(\kappa H)^2} \simeq 0,
\end{equation}
and standard manipulations then yield
\begin{equation}\label{eq:psi_0_thin_modified}
\psi_0 \simeq \ln\left[ \frac{2+(\rho \kappa H)^{2}+\rho \kappa H \sqrt{ 4 + (\kappa H)^{2}(1 + \rho^2)}}{2+\rho (\kappa H)^{2} }  \right]
\end{equation}

For sufficiently large $\rho$, this may be reexpressed as
	\begin{equation}\label{eq:psi_0_thin_modified2}
	\psi_0 \simeq \ln\left[ \frac{\left(\dfrac{\rho \kappa H}{2} + \sqrt{ 1 + \left(\dfrac{\rho \kappa H}{2}\right)^{2}}\right)^2}{1+ \dfrac{\rho (\kappa H)^{2}}{2}} \right],
	\end{equation}
which \textbf{is equivalent to}
\begin{equation}\label{eq:psi_0_thin_modified3}
\psi_0 \simeq 2 \arsinh\left(\dfrac{\rho \kappa H}{2}\right)  - \ln \left(1 +\dfrac{\rho (\kappa H)^2}{2}\right).
\end{equation}

For small $\rho \kappa H$, the $\psi$-profile is almost constant throughout the film,
\begin{equation}\label{eq:small_rho_kappa_H}
    \psi_0 \simeq  \psi_s \simeq \rho \kappa H,
\end{equation}
as follows directly from \eqref{eq:psi_0_thin_modified}.

Finally, the inner $\psi$-profile of a thin film is given by
\begin{equation}\label{eq:phi_in_thin1}
\psi_{i}(z)\simeq \psi_s-\frac{\left[\sinh \psi_0-\rho\right]}{2}\kappa^2(H^2-z^2),
\end{equation}
where $\psi_s$ and $\psi_0$ are described by Eqs.\eqref{eq:psi_s_thick} and \eqref{eq:psi_0_thin_modified3}.


The startling conclusion from analysis of Eq.\eqref{eq:rho_of_psi} is that it is also valid for $\kappa H \gg 1$, i.e. thick films, where the inner region far from the interface may be modeled as electroneutral. Indeed, in this case the last term on its left-hand side becomes very small compared with the first two, and we obtain a well-known  for thick films result~\cite{ohshima.h:1985}
\begin{equation}\label{eq:phi_0_thick}
\psi_0=\arsinh(\rho),
\end{equation}
which leads to $p_0 = \sqrt{1+\rho^2}$. The potential of an electroneutral area of a thick film is usually referred to as the Donnan potential, $\psi_D$, so that Eq.\eqref{eq:phi_0_thick} relates $\psi_D$ with $\rho$. Note that in this limit Eq.\eqref{eq:psi_s_thick} can be transformed to
\begin{equation}\label{eq:psi_s_thick1}
\psi_s = \psi_0 + \displaystyle\frac{1-\sqrt{1+\rho^2}}{\rho}
\end{equation}

Near the surface electrolyte ions screen volume charges of the coating  only partly, and an inner diffuse layer is formed. The inverse inner screening length can be found as
\begin{equation}
\kappa_{i} = \kappa (\cosh{\psi_0})^{1/2} \equiv \kappa \sqrt{p_0}
\end{equation}
From  Eq.\eqref{eq:phi_0_thick} it then follows that $\kappa_{i} \simeq \kappa (1+\rho^{2})^{1/4}$, which indicates that when $\rho\ll 1$, a sensible approximation should be $\kappa_{i} \simeq \kappa$. However, when $\rho\gg1$, $\kappa_{i} \simeq \kappa \sqrt{\rho}$, and the criterion of a thick film can be relaxed to $\kappa H \sqrt{\rho}  \gg 1$.

Since the thick film `behaves' as an electrolyte solution of the inverse Debye length $\kappa_{i}$, to obtain the exact equation for $\psi_i$  it is enough to simply change the variables in Eq.\eqref{eq:PB_out3}. Namely, substitution of $z$ by $-z$, $\psi_s$ by $\psi_{0} - \psi_{s}$, and $\psi_o$ by $\psi_{0} - \psi_{i}$ would immediately give

\begin{equation}\label{eq:new}
\psi_{0}-\psi_{i} = 4 \artanh \left[ \gamma_{i} e^{-\kappa_{i}(H-z)}\right],
\end{equation}
 where
$  \gamma_{i} = \tanh{\dfrac{\psi_{0}-\psi_{s}}{4}}$. Using Eqs.\eqref{eq:phi_0_thick} and \eqref{eq:psi_s_thick1} we obtain $\gamma_{i} = \tanh\left({\dfrac{\sqrt{1 + \rho^2} - 1}{4\rho}}\right)$, which reduces to $\gamma_{i} \simeq \rho/8$ if $\rho\ll 1$, and $\gamma_{i} \simeq \dfrac{1}{4}\left(1-\dfrac{1}{\rho}\right)$ when $\rho\gg 1$. This implies that $\gamma_{i}$ is always smaller than 1/4. For such a small $\gamma_{i}$, the inner potential can be expanded about $\psi_0$, and to first order in $\psi_0 - \psi_s$ we obtain
\begin{equation}\label{eq:phi_in_thick2}
\psi_{i}(z) \simeq \psi_0+(\psi_s-\psi_0)e^{\kappa_i(z-H)}
\end{equation}
This derivation differs from conventional arguments, which assume low volume charge density~\cite{ohshima.h:1985}. Our treatment clarifies that Eq.\eqref{eq:phi_in_thick2} constitutes a sensible approximation for $\psi_i$ of a thick film of any $\rho$.

\begin{figure}[t]
\begin{center}
\includegraphics[width=1\columnwidth]{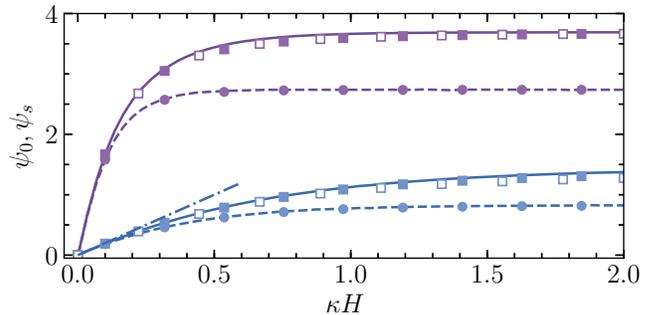}
\caption{Potentials at wall (solid lines) and surface (dashed)  as a function of $\kappa H$ computed for fixed $\rho = 20$ (upper set of curves) and $\rho = 2$ (lower curves). Filled squares illustrate calculations from Eq.\eqref{eq:psi_0_thin_modified}, circles are then obtained using  Eq.\eqref{eq:psi_s_thick}. Open squares show results obtained using Eq.\eqref{eq:psi_0_thin_modified3}. Dash-dotted line is calculated from Eq.\eqref{eq:small_rho_kappa_H}. }\label{fig:Fig3}
\end{center}
\end{figure}

In order to assess the validity of the above approach, we employ numerical simulations. We perform a numerical resolution of a multi-point boundary value problem for the nonlinear Poisson-Boltzmann differential equation Eq.~\eqref{Eq:PB_io} with prescribed boundary conditions, using the numerical approach based on the collocation method developed by \citet{bader.g:1987}. 


In Fig.~\ref{fig:Fig2} we plot $\psi (z/H)$  computed for two different values of $\kappa H$ that are close to limits of thick and thin films, and a large fixed $\rho$. The form of the $\psi$-profile depends on $\kappa H$. For $\kappa H = 3$ the inner potential shows a distinct plateau indicating that the intrinsic charge of the film is completely screened by electrolyte ions, i.e. global electroneutrality. The plateau potential is equal to $\psi_D$. However, when $\kappa H = 0.3$, there is no electroneutral region inside the film, and the potential at wall, $\psi_0$, is much smaller than $\psi_D$. Also included in Fig.~\ref{fig:Fig2} are theoretical results obtained from Eqs.\eqref{eq:PB_out3}, \eqref{eq:phi_in_thin1}, \eqref{eq:new}, and \eqref{eq:phi_in_thick2} and we conclude that in relevant areas they are in excellent agreement with numerical data.

\begin{figure}[t]
\begin{center}
\includegraphics[width=1\columnwidth]{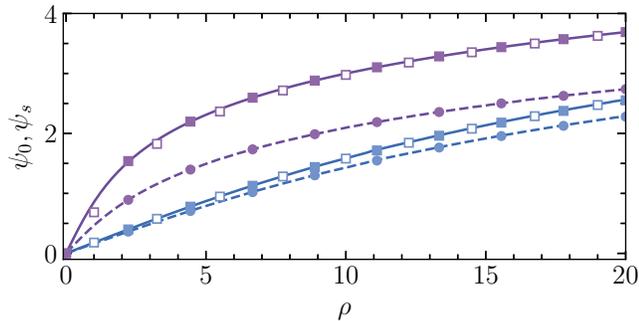}
\caption{Potentials at wall (solid lines) and surface (dashed)  as a function of $\rho$ computed for fixed $\kappa H = 3$ (upper set of curves) and $\kappa H = 0.2$ (lower curves). Filled squares illustrate calculations from Eq.\eqref{eq:psi_0_thin_modified}, circles are then obtained using  Eq.\eqref{eq:psi_s_thick}. Open squares show predictions of Eq.\eqref{eq:psi_0_thin_modified3}.}\label{fig:Fig4}
\end{center}
\end{figure}

It is tempting to speculate that Eq.\eqref{eq:psi_0_thin_modified} will be applicable for any $\kappa H$, and that a more elegant result, Eq.\eqref{eq:psi_0_thin_modified3}, can be used provided $\rho$ large enough. Clearly,  Eq.\eqref{eq:rho_of_psi} could become less accurate for intermediate $\kappa H$, and it is of considerable interest to determine its regime of validity.
To test ansatz Eq.\eqref{eq:rho_of_psi}, numerical and theoretical $\psi_0$ and $\psi_s$ have been calculated as a function of $\kappa H$  for different values of $\rho$. Specimen results are plotted in Fig.~\ref{fig:Fig3} confirming the validity of Eqs.\eqref{eq:psi_0_thin_modified} and \eqref{eq:psi_0_thin_modified3}  for all $\kappa H$. As expected, Eq.\eqref{eq:small_rho_kappa_H}, which can also be obtained using the linear theory~\cite{ohshima.h:1985}, is valid only when $\rho \kappa H$ is very small and significantly overestimates potentials, which saturate at some $\kappa H$, in other cases. The charge density dependence of $\psi_0$ and $\psi_s$  is also of interest. Fig.~\ref{fig:Fig4} illustrates the weak growth of both potentials with $\rho$, and that $\psi_0 - \psi_s \simeq 1$ as $\rho \kappa H$ is increased. We again conclude that Eq.\eqref{eq:psi_0_thin_modified} fits accurately the numerical data. So does  \eqref{eq:psi_0_thin_modified3}, except for $\rho \leq 1$, where some very small discrepancy is observed. Below we use
Eq.\eqref{eq:psi_0_thin_modified3} for all calculations, by omitting  a discussion of the accuracy of our theory.


\subsection{Ion concentrations}

\begin{figure}[t]
	\begin{center}
	\includegraphics[width=1\columnwidth]{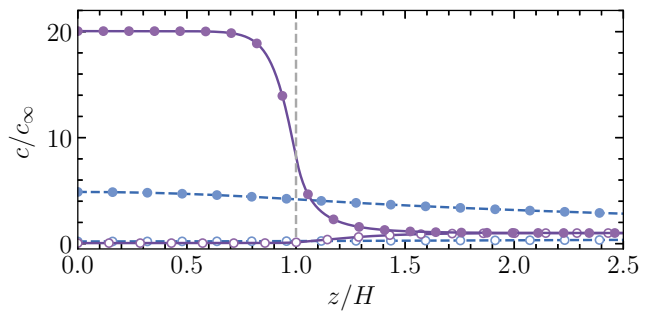}
        \caption{Ion concentration profiles computed at $\rho = 10$ using $\kappa H = 0.2$ (dashed curves) and $5$ (solid curves). Local concentrations of anions (filled circles) and cations (open circles) are calculated using Eq.\eqref{eq:PB_out3} when $z/H \geq  1$, and from Eqs.\eqref{eq:phi_in_thin1} and Eq.\eqref{eq:new} when $z/H \leq  1$.  }
		\label{fig:concentrations}
	\end{center}
\end{figure}

The problem we address here is the calculation of the profile of a cloud of counter-ions forming a diffuse electrostatic layer close to a planar surface of a porous film. Since ions obey Boltzmann distribution, their local concentration $c_{\pm}/c_{\infty} = \exp{(\mp \psi})$ are determined solely by the $\psi$-profile calculated above.
Representative concentration  profiles   computed for highly charged porous films of $\rho = 10$ and two different values of $\kappa H$  are  shown  in Fig.~\ref{fig:concentrations}. We see that in the inner region anions are    significantly enriched, and cations are depleted. The degree of this enrichment and depletion depends on the values of $\rho$ and $\kappa H$. At the given $\rho$ the degree of enrichment is ca.20 for a thick film of $k H = 5$, but it is a few times smaller when $k H = 0.2$. We also stress that for a chosen value of $\rho$ inner concentrations of cations practically vanish for both $\kappa H$.

\begin{figure}[t]
\begin{center}
\includegraphics[width=1\columnwidth]{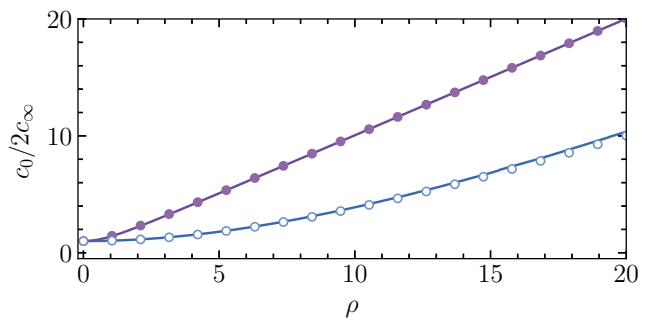}
\caption{Ion-enrichment at the wall, $c_0/2c_{\infty} = \cosh \psi_0$, vs. $\rho$ computed using $\kappa H = 3$ (upper curve) and $0.3$ (lower curve). Filled and open circles are obtained from Eqs.\eqref{eq:phi_0_thick} and \eqref{eq:psi_0_thin_modified}.}\label{Fig:c0_rho}
\end{center}
\end{figure}

 To boost absorption of ions, coatings of larger $\rho$  can be used, as illustrated in Fig.~\ref{Fig:c0_rho}, where the total concentration of ions at the wall, $c_0 = c_+(0) + c_-(0),$ scaled by the sum of  anion and cation concentrations at infinity, $2 c_{\infty}$, is plotted as a function of $\rho$. We note that for a thick film $c_{0} \equiv p_0$, as follows from Eq.~\eqref{eq:phi_0_thick}, so that it grows linearly with $\rho$, when it becomes large enough.
 Eq.\eqref{eq:phi_0_thick} describes perfectly numerical data for $\kappa H = 3$, and it is clear that this curve corresponds to an upper attainable value of ion-enrichment at the wall. In other words, the absorption capacity cannot be further improved by making the coating thicker. However, the reduction in $H$ could significantly reduce the concentration of absorbed anions, especially at $\rho \leq 10$.


\section{Electroosmotic Velocity and zeta potential}

Another relevant problem is an electro-osmotic flow of a solvent of the dynamic viscosity $\eta$ in an applied tangential electric field, $E$. The origin of the electro-osmotic flow is traditionally attributed to diffuse layers. Below we show that for porous coatings the electroosmosis is defined both by diffuse layers and the
absorbed ions, and that \textbf{the second mechanism is} responsible for a flow amplification and dominates at high salt.

We are now about to relate the dimensionless velocity, $v(z) = \dfrac{4\pi\ell_{B}\eta}{eE} V(z)$, of such a flow to $\psi_0$ and $\psi_s$. We assume weak field, so that in steady state $\psi(z)$ is independent of the fluid flow. Note that for our planar geometry the concentration gradients at every
location are perpendicular to the direction of the flow, it is therefore legitimate to neglect advection. Therefore, the liquid flow satisfies the generalized  Stokes equation
\begin{equation}\label{eq:Stokes}
v'' - \mathcal{K}^2 v \Theta(H-z)   = \psi'' +  \kappa^2 \rho \Theta(H-z),
\end{equation}
where $\mathcal{K}$ is the inverse Brinkman length.
 At the wall we apply a classical no-slip condition, $v_0 = v (0) = 0$, and far from the surface $v'_{z\rightarrow\infty} = 0$.
 We consider the limits of small flow extension into the porous medium, $\mathcal{K} H \to \infty$,  and of $\mathcal{K} H \to 0$, where an additional dissipation in the porous film can be  neglected, to obtain bounds on the electro-osmotic velocity that constrain its attainable value.

From analysis of Eq.\eqref{eq:Stokes} it follows that the outer $v$-profile and velocity in the bulk are given by
\begin{equation}\label{eq:v_inf}
 v_{o}(z) = v_{\infty} + \psi_{o}(z),  \, v_{\infty} = v_s - \psi_{s} = - \zeta,
\end{equation}
where $\psi_{o}$ is defined by Eq.\eqref{eq:PB_out3}, $v_s = v(H)$ is the liquid velocity at surface, below we refer it to as slip velocity, and $\zeta = e Z / (k_B T) $. Eqs.(\ref{eq:v_inf}) indicate that enhanced electro-osmotic mobility can be a consequence of large equilibrium $\psi_s$, as well as of large $v_s$ that depends on the hydrodynamic permeability of the coating. The amplification of the electro-osmotic flow (compared to the no-slip case with the same $\psi_s$) can be expressed as
\begin{equation}\label{eq:amplification}
    \frac{\zeta}{\psi_s} = 1-  \frac{v_s}{\psi_s}
\end{equation}
The problem, thus, reduces to calculation of $v_s$. Below we provide analytical results together with exact numerical calculations.

\begin{figure}[t]
\begin{center}
\includegraphics[width=1\columnwidth]{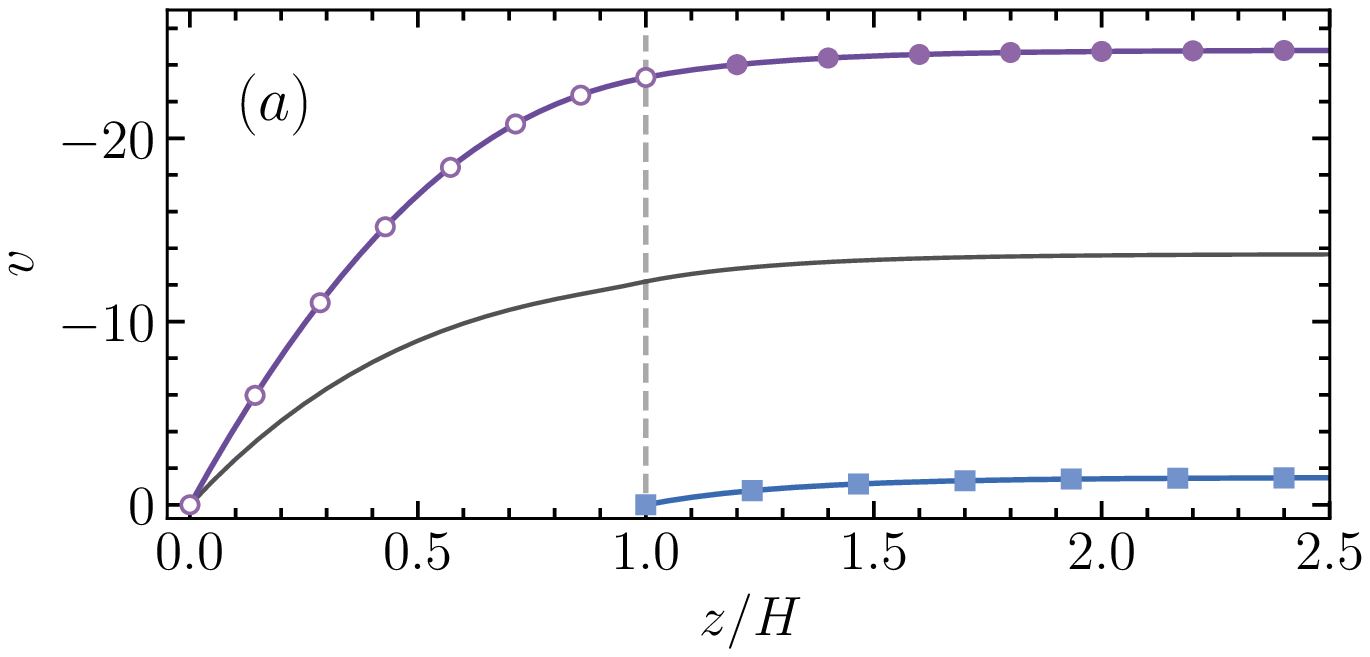}
\includegraphics[width=1\columnwidth]{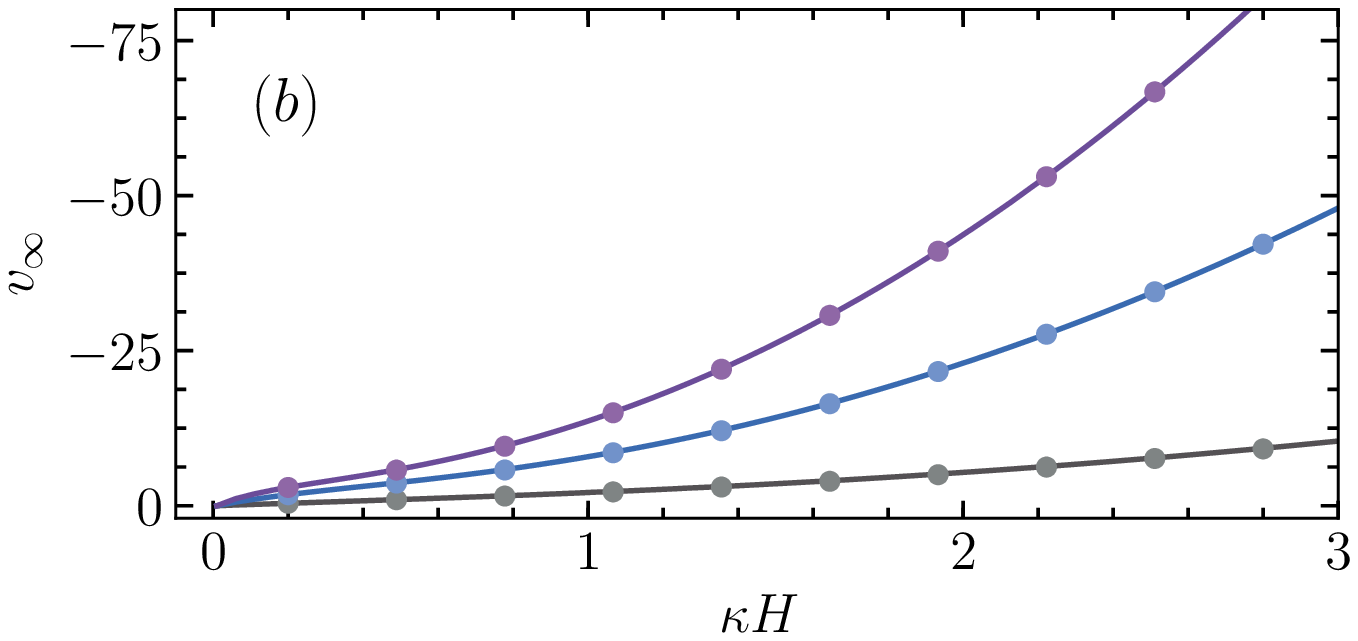}
\caption{(a) Electroosmotic velocity profiles computed using $\kappa H = 3$, $\rho = 5$, and $\mathcal{K} H = 0 $, $1.5$,  and $\infty$ (solid curves from top to bottom). Filled and open symbols indicate calculations from Eqs.~\eqref{eq:v_inf} and Eq.~\eqref{eq:l_inf_in}.
 Squares are obtained using $v_s = 0$, circles correspond to $v_s = \psi_s + v_{\infty}$, where $v_{\infty}$ is given by Eq.~\eqref{eq:l_inf_v_inf}; (b) Computed upper bounds ($\mathcal{K} H \to 0$) on $v_{\infty}$ vs. $\kappa H$  (solid curves). From top to bottom $\rho = 20, 10,$ and $2$. Circles show results of calculations from  Eq.\eqref{eq:l_inf_v_inf}.}\label{Fig:eo}
\end{center}
\end{figure}

If we suppose $\mathcal{K} H \to \infty$, the slip velocity nearly vanishes, $\psi_{s} \simeq \zeta$, and $v_{\infty} \simeq - \psi_{s}$, which is equivalent to the Smoluchowsky result. Fig.~\ref{Fig:eo}(a) includes typical numerical and theoretical $v$-profiles calculated for this case using $\kappa H = 3$ and $\rho = 5$ that leads to $\psi_s \simeq 1.5$.
When $\mathcal{K} H \to 0$, integrating Eq.\eqref{eq:Stokes} twice, and imposing the continuity of $v$ and $v'$ at $z=H$, we find
\begin{equation}\label{eq:l_inf_in}
    v_{i} \simeq (\psi_{i} - \psi_{0}) - \rho \kappa^2 \left( Hz - \frac{z^2}{2} \right),
\end{equation}
where for films of any thickness $\psi_{0}$ is given by Eq.~\eqref{eq:psi_0_thin_modified}. The first term reflects the reduction of the
potential in the porous coating. The second term is associated with a body force $\rho \kappa^2$ that drives the inner flow by acting on the accumulated mobile ions. This contribution resembles  the usual no-slip parabolic Poiseuille flow. It follows from Eq.\eqref{eq:l_inf_in} that
$ - v_s \simeq \psi_{0} - \psi_{s} + \dfrac{\rho (\kappa H)^{2}}{2}$. Since $\psi_{0} - \psi_{s} \leq 1$, the second term should dominate  even at moderate $\rho$.

The outer velocity $v_ o$ is given by Eqs.\eqref{eq:v_inf} with
\begin{equation}\label{eq:l_inf_v_inf}
v_{\infty} \simeq - \left(\psi_{0} + \dfrac{\rho (\kappa H)^2}{2}\right) \simeq -\zeta
\end{equation}
The $v$-profile for this case is also shown in Fig.~\ref{Fig:eo}(a). It turns out that even at moderate $\rho$ and $\kappa H$ one can induce  significantly enhanced $v_{\infty}$, which is associated with the emergence of a large slip velocity, $v_s$. Also included in Fig.~\ref{Fig:eo}(a) is the curve computed using $\mathcal{K} H = 1.5$, which is located between two limiting cases and demonstrates quite large $v_s$.

We now verify Eq.\eqref{eq:l_inf_v_inf} and plot theoretical $v_{\infty}$ vs. $\kappa H$ in Fig.~\ref{Fig:eo}(b) together with numerical data. Upon increasing $\kappa H$ at fixed $\rho$ the amplitude of $v_{\infty}$ grows nonlinearly, and in the case of $\rho \gg 1$ becomes  several tens of times faster compared to a no-slip case, even at moderate $\kappa H$. It is tempting to speculate that one can further amplify $v_{\infty}$ making porous film thicker. However, when the film becomes thick enough, the condition $\mathcal{K} H \to 0$ violates, and Eq.\eqref{eq:l_inf_v_inf} is no longer valid.

This implies that mobile ions absorbed within the porous layer actively participate in the flow-driving mechanism by reacting to the field. The porous film acts as a charged immobile surface layer with absorbed mobile ions of the opposite sign, but note the difference from a known example of mobile surface charges at slippery wall~\cite{maduar.sr:2015}. In the latter case, slippage is of a hydrodynamic origin and mobile surface charges induce a backward flow,  reducing the amplification of electro-osmotic flow caused by hydrodynamic slip. By contrast, in the current work  an inner solvent flow induces a forward flow and $v_s$ itself. However, similarly to hydrophobic electrokinetics~\cite{joly2004,maduar.sr:2015}, our large $\zeta$ no longer reflects the sole $\psi_s$. Finally, we would like to stress that a massive amplification of electro-osmotic flow that can be achieved  near porous surfaces is of the same order of that at charged super-hydrophobic surfaces~\cite{belyaev.av:2011a}, and such a fast flow is in agreement   with recent observations~\cite{feldmann.d:2020}.

\begin{figure}[t]
	\begin{center}
		\includegraphics[width=1\columnwidth]{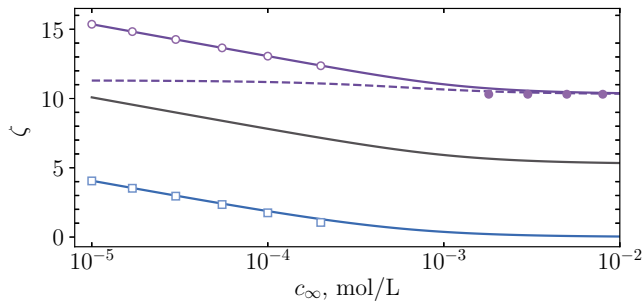}
		\caption{$\zeta$ as a function of $c_{\infty}$ computed for a film of $H = 50$ nm, $\varrho = 150$ kC/m$^3$, using $\mathcal{K} H = 0 $, $1.5$, and $\infty$ (solid curves from top to bottom). Dashed curve shows $|v_s|$ at $\mathcal{K} H = 0 $.
		Squares, open and filled circles plot results of calculations from Eqs.\eqref{eq:zeta_low}, \eqref{eq:log_plateau}, and \eqref{eq:plateau}.  }\label{Fig:zeta_c}
	\end{center}
\end{figure}

 So far we have considered the potentials and electro-osmotic
velocity at fixed dimensionless $\rho$ and $\kappa H$. Additional insight into the problem can be gleaned by calculating an upper bound on $\zeta$ as a function of $c_{\infty} \propto \kappa^{2}$ at fixed $H$ and $\varrho$. Let us now keep fixed $H = 50$ nm and $\varrho = 150$ kC/m$^3$, usually referred to as `moderate'~\cite{duval.j:2009}, and vary $c_{\infty}$ from $10^{-5}$ to $10^{-2}$ mol/L.
Upon increasing $c_{\infty}$ in this range, $\kappa H$ is increased from about 0.5 to 16, and $\rho$ is reduced from about 78 down to 0.1. Therefore, for a given film the required regimes (of thin and thick films, or highly and weakly charged coatings) can be  tuned simply by adjusting the concentration of salt.

  The bounds on $\zeta$ are shown by lower and  upper curves in Fig.~\ref{Fig:zeta_c}.
If $\mathcal{K} H \to \infty$, $\zeta \simeq \psi_s$ decays from ca. 5 ($Z \simeq 125$ mV) practically to zero as $c_{\infty}$ increases, leading to a suppression of a flow. In dilute solutions the lower bound is given by
\begin{equation}\label{eq:zeta_low}
     \zeta \simeq \ln{\left(\frac{\varrho}{ec_{\infty}}\right)}-1,
\end{equation}
which perfectly fits numerical data.
When $\mathcal{K} H \to 0$, $\zeta$ becomes much larger. Thus, with our smallest concentration $\zeta \simeq 15$ (or $Z \simeq 375$ mV).
In a dilute solution:
\begin{equation}\label{eq:log_plateau}
    \zeta \simeq \ln{\left(\frac{\varrho}{ec_{\infty}}\right)}  + \dfrac{2\pi\ell_{B}\varrho H^2}{e},
\end{equation}
where first term is associated with the potential at the wall. Note that such a logarithmic decay  fits well the obtained for real porous materials data~\cite{feldmann.d:2020}.
For concentrated solutions a large, and independent on salt, zeta-potential is observed. The computed slip velocity,  also included in Fig.~\ref{Fig:zeta_c}, indicates that this occurs when $\zeta \simeq -v_s$ and $\psi_s \simeq 0$. Thus, a large zeta-potential emerges solely due to  a forward electro-osmotic flow inside a porous film. It is easy to show that it is given by
\begin{equation}\label{eq:plateau}
    \zeta  \simeq \dfrac{ 2 \pi \ell_B   \varrho H^2}{e},
\end{equation}
that clarifies the status of second term in Eq.\eqref{eq:log_plateau}.
This result is relevant for the understanding zeta-potential measurements with `hairy' surfaces, where it remains finite at even high salt concentrations~\cite{donath.e:1979,garg.a:2016}. We recall that Eqs.\eqref{eq:zeta_low} and \eqref{eq:log_plateau} represent the lower and upper bounds for $\zeta$ attained at limiting values of $\mathcal{K} H$. Any finite $\mathcal{K} H$ would lead to $\zeta$ confined between the above bounds, as seen in Fig.~\ref{Fig:zeta_c}, where we use $\mathcal{K} H = 1.5$.

\section{Conclusion}

In summary, our nonlinear analytic model provides  considerable insight into the electro-osmotic equilibria and flows in the presence of porous coatings. It describes absorption capacity of porous films, which is, in turn, responsible for an enhanced electro-osmotic flow. The bounds on zeta-potential we have derived  can guide the design of coatings to amplify  electrokinetic phenomena.

\appendix






\begin{acknowledgments}

 We thank E.S.Asmolov for helpful  discussions. This work was supported by the Ministry of Science and Higher Education of the Russian Federation and by the German Research Foundation (grant Vi 243/4-2) within the SPP 1726.
\end{acknowledgments}


\bibliography{Bibliography}

\begin{thebibliography}{24}%
\makeatletter
\providecommand \@ifxundefined [1]{%
 \@ifx{#1\undefined}
}%
\providecommand \@ifnum [1]{%
 \ifnum #1\expandafter \@firstoftwo
 \else \expandafter \@secondoftwo
 \fi
}%
\providecommand \@ifx [1]{%
 \ifx #1\expandafter \@firstoftwo
 \else \expandafter \@secondoftwo
 \fi
}%
\providecommand \natexlab [1]{#1}%
\providecommand \enquote  [1]{``#1''}%
\providecommand \bibnamefont  [1]{#1}%
\providecommand \bibfnamefont [1]{#1}%
\providecommand \citenamefont [1]{#1}%
\providecommand \href@noop [0]{\@secondoftwo}%
\providecommand \href [0]{\begingroup \@sanitize@url \@href}%
\providecommand \@href[1]{\@@startlink{#1}\@@href}%
\providecommand \@@href[1]{\endgroup#1\@@endlink}%
\providecommand \@sanitize@url [0]{\catcode `\\12\catcode `\$12\catcode
  `\&12\catcode `\#12\catcode `\^12\catcode `\_12\catcode `\%12\relax}%
\providecommand \@@startlink[1]{}%
\providecommand \@@endlink[0]{}%
\providecommand \url  [0]{\begingroup\@sanitize@url \@url }%
\providecommand \@url [1]{\endgroup\@href {#1}{\urlprefix }}%
\providecommand \urlprefix  [0]{URL }%
\providecommand \Eprint [0]{\href }%
\providecommand \doibase [0]{http://dx.doi.org/}%
\providecommand \selectlanguage [0]{\@gobble}%
\providecommand \bibinfo  [0]{\@secondoftwo}%
\providecommand \bibfield  [0]{\@secondoftwo}%
\providecommand \translation [1]{[#1]}%
\providecommand \BibitemOpen [0]{}%
\providecommand \bibitemStop [0]{}%
\providecommand \bibitemNoStop [0]{.\EOS\space}%
\providecommand \EOS [0]{\spacefactor3000\relax}%
\providecommand \BibitemShut  [1]{\csname bibitem#1\endcsname}%
\let\auto@bib@innerbib\@empty
\bibitem [{\citenamefont {Porada}\ \emph {et~al.}(2012)\citenamefont {Porada},
  \citenamefont {Weinstein}, \citenamefont {Dash}, \citenamefont {van~der Wal},
  \citenamefont {Bryjak}, \citenamefont {Gogotsi},\ and\ \citenamefont
  {Biesheuvel}}]{porada.s:2012}%
  \BibitemOpen
  \bibfield  {author} {\bibinfo {author} {\bibfnamefont {S.}~\bibnamefont
  {Porada}}, \bibinfo {author} {\bibfnamefont {L.}~\bibnamefont {Weinstein}},
  \bibinfo {author} {\bibfnamefont {R.}~\bibnamefont {Dash}}, \bibinfo {author}
  {\bibfnamefont {A.}~\bibnamefont {van~der Wal}}, \bibinfo {author}
  {\bibfnamefont {M.}~\bibnamefont {Bryjak}}, \bibinfo {author} {\bibfnamefont
  {Y.}~\bibnamefont {Gogotsi}}, \ and\ \bibinfo {author} {\bibfnamefont
  {P.~M.}\ \bibnamefont {Biesheuvel}},\ }\bibfield  {title} {\enquote {\bibinfo
  {title} {Water desalination using capacitive deionization with microporous
  carbon electrodes},}\ }\href@noop {} {\bibfield  {journal} {\bibinfo
  {journal} {ACS Appl. Mater. Interfaces}\ }\textbf {\bibinfo {volume} {4}},\
  \bibinfo {pages} {1194--1199} (\bibinfo {year} {2012})}\BibitemShut {NoStop}%
\bibitem [{\citenamefont {Stamatialis}\ \emph {et~al.}(2008)\citenamefont
  {Stamatialis}, \citenamefont {Papenburg}, \citenamefont {Gironas},
  \citenamefont {Saiful}, \citenamefont {Bettahalli}, \citenamefont
  {Schmitmeier},\ and\ \citenamefont {Wessling}}]{stamatialis.df:2008}%
  \BibitemOpen
  \bibfield  {author} {\bibinfo {author} {\bibfnamefont {D.}~\bibnamefont
  {Stamatialis}}, \bibinfo {author} {\bibfnamefont {B.J.}\ \bibnamefont
  {Papenburg}}, \bibinfo {author} {\bibfnamefont {M.}~\bibnamefont {Gironas}},
  \bibinfo {author} {\bibfnamefont {S.}~\bibnamefont {Saiful}}, \bibinfo
  {author} {\bibfnamefont {S.N.M.}\ \bibnamefont {Bettahalli}}, \bibinfo
  {author} {\bibfnamefont {S.}~\bibnamefont {Schmitmeier}}, \ and\ \bibinfo
  {author} {\bibfnamefont {M.}~\bibnamefont {Wessling}},\ }\bibfield  {title}
  {\enquote {\bibinfo {title} {Medical applications of membranes: Drug
  delivery, artificial organs and tissue engineering},}\ }\href {\doibase
  https://doi.org/10.1016/j.memsci.2007.09.059} {\bibfield  {journal} {\bibinfo
   {journal} {J. Membrane Sci.}\ }\textbf {\bibinfo {volume} {308}},\ \bibinfo
  {pages} {1 -- 34} (\bibinfo {year} {2008})}\BibitemShut {NoStop}%
\bibitem [{\citenamefont {Chung}\ \emph {et~al.}(2018)\citenamefont {Chung},
  \citenamefont {Mireles}, \citenamefont {Kwarta},\ and\ \citenamefont
  {Gaborski}}]{chung.hh:2018}%
  \BibitemOpen
  \bibfield  {author} {\bibinfo {author} {\bibfnamefont {H.~H.}\ \bibnamefont
  {Chung}}, \bibinfo {author} {\bibfnamefont {M.}~\bibnamefont {Mireles}},
  \bibinfo {author} {\bibfnamefont {B.~J.}\ \bibnamefont {Kwarta}}, \ and\
  \bibinfo {author} {\bibfnamefont {T.~R.}\ \bibnamefont {Gaborski}},\
  }\bibfield  {title} {\enquote {\bibinfo {title} {Use of porous membranes in
  tissue barrier and co-culture models},}\ }\href@noop {} {\bibfield  {journal}
  {\bibinfo  {journal} {Lab on a Chip}\ }\textbf {\bibinfo {volume} {18}},\
  \bibinfo {pages} {1671--1689} (\bibinfo {year} {2018})}\BibitemShut {NoStop}%
\bibitem [{\citenamefont {Biesheuvel}\ \emph {et~al.}(2011)\citenamefont
  {Biesheuvel}, \citenamefont {Fu},\ and\ \citenamefont
  {Bazant}}]{biesheuvel.pm:2011}%
  \BibitemOpen
  \bibfield  {author} {\bibinfo {author} {\bibfnamefont {P.~M.}\ \bibnamefont
  {Biesheuvel}}, \bibinfo {author} {\bibfnamefont {Y.}~\bibnamefont {Fu}}, \
  and\ \bibinfo {author} {\bibfnamefont {M.~Z.}\ \bibnamefont {Bazant}},\
  }\bibfield  {title} {\enquote {\bibinfo {title} {Diffuse charge and faradaic
  reactions in porous electrodes},}\ }\href@noop {} {\bibfield  {journal}
  {\bibinfo  {journal} {Phys. Rev. E}\ }\textbf {\bibinfo {volume} {83}},\
  \bibinfo {pages} {061507} (\bibinfo {year} {2011})}\BibitemShut {NoStop}%
\bibitem [{\citenamefont {Feldmann}\ \emph {et~al.}(2020)\citenamefont
  {Feldmann}, \citenamefont {Arya}, \citenamefont {Molotilin}, \citenamefont
  {Lomadze}, \citenamefont {Kopyshev}, \citenamefont {Vinogradova},\ and\
  \citenamefont {Santer}}]{feldmann.d:2020}%
  \BibitemOpen
  \bibfield  {author} {\bibinfo {author} {\bibfnamefont {D.}~\bibnamefont
  {Feldmann}}, \bibinfo {author} {\bibfnamefont {P.}~\bibnamefont {Arya}},
  \bibinfo {author} {\bibfnamefont {T.~Y.}\ \bibnamefont {Molotilin}}, \bibinfo
  {author} {\bibfnamefont {N.}~\bibnamefont {Lomadze}}, \bibinfo {author}
  {\bibfnamefont {A.}~\bibnamefont {Kopyshev}}, \bibinfo {author}
  {\bibfnamefont {O.~I.}\ \bibnamefont {Vinogradova}}, \ and\ \bibinfo {author}
  {\bibfnamefont {S.~A.}\ \bibnamefont {Santer}},\ }\bibfield  {title}
  {\enquote {\bibinfo {title} {Extremely long-range light-driven repulsion of
  porous microparticles},}\ }\href@noop {} {\bibfield  {journal} {\bibinfo
  {journal} {Langmuir}\ }\textbf {\bibinfo {volume} {36}},\ \bibinfo {pages}
  {6994--7004} (\bibinfo {year} {2020})}\BibitemShut {NoStop}%
\bibitem [{\citenamefont {Donath}\ and\ \citenamefont
  {Pastushenko}(1979)}]{donath.e:1979}%
  \BibitemOpen
  \bibfield  {author} {\bibinfo {author} {\bibfnamefont {E.}~\bibnamefont
  {Donath}}\ and\ \bibinfo {author} {\bibfnamefont {V.}~\bibnamefont
  {Pastushenko}},\ }\bibfield  {title} {\enquote {\bibinfo {title}
  {Eletrophoretical study of cell surface properties. {T}he influence of the
  surface coat on the electric potential distribution and on general
  electrokinetic properties of animal cells},}\ }\href@noop {} {\bibfield
  {journal} {\bibinfo  {journal} {Bioelectrochem. Bioenergetics}\ }\textbf
  {\bibinfo {volume} {6}},\ \bibinfo {pages} {543--554} (\bibinfo {year}
  {1979})}\BibitemShut {NoStop}%
\bibitem [{\citenamefont {Ohshima}\ and\ \citenamefont
  {Ohki}(1985)}]{ohshima.h:1985}%
  \BibitemOpen
  \bibfield  {author} {\bibinfo {author} {\bibfnamefont {H.}~\bibnamefont
  {Ohshima}}\ and\ \bibinfo {author} {\bibfnamefont {S.}~\bibnamefont {Ohki}},\
  }\bibfield  {title} {\enquote {\bibinfo {title} {Donnan potential and surface
  potential of a charged membrane},}\ }\href@noop {} {\bibfield  {journal}
  {\bibinfo  {journal} {Biophys. J.}\ }\textbf {\bibinfo {volume} {47}},\
  \bibinfo {pages} {673--678} (\bibinfo {year} {1985})}\BibitemShut {NoStop}%
\bibitem [{\citenamefont {Ohshima}\ and\ \citenamefont
  {Kondo}(1990)}]{ohshima.h:1990a}%
  \BibitemOpen
  \bibfield  {author} {\bibinfo {author} {\bibfnamefont {H.}~\bibnamefont
  {Ohshima}}\ and\ \bibinfo {author} {\bibfnamefont {T.}~\bibnamefont
  {Kondo}},\ }\bibfield  {title} {\enquote {\bibinfo {title} {Electrokinetic
  flow between two parallel plates with surface charge layers: electro-osmosis
  and streaming potential},}\ }\href@noop {} {\bibfield  {journal} {\bibinfo
  {journal} {J. Colloid Interface Sci.}\ }\textbf {\bibinfo {volume} {135}},\
  \bibinfo {pages} {443--448} (\bibinfo {year} {1990})}\BibitemShut {NoStop}%
\bibitem [{\citenamefont {Chanda}\ \emph {et~al.}(2014)\citenamefont {Chanda},
  \citenamefont {Sinha},\ and\ \citenamefont {Das}}]{chanda.s:2014}%
  \BibitemOpen
  \bibfield  {author} {\bibinfo {author} {\bibfnamefont {S.}~\bibnamefont
  {Chanda}}, \bibinfo {author} {\bibfnamefont {S.}~\bibnamefont {Sinha}}, \
  and\ \bibinfo {author} {\bibfnamefont {S.}~\bibnamefont {Das}},\ }\bibfield
  {title} {\enquote {\bibinfo {title} {Streaming potential and electroviscous
  effects in soft nanochannels: towards designing more efficient nanofluidic
  electrochemomechanical energy converters},}\ }\href@noop {} {\bibfield
  {journal} {\bibinfo  {journal} {Soft Matter}\ }\textbf {\bibinfo {volume}
  {10}},\ \bibinfo {pages} {7558--7568} (\bibinfo {year} {2014})}\BibitemShut
  {NoStop}%
\bibitem [{\citenamefont {Duval}(2005)}]{duval.jfl:2005}%
  \BibitemOpen
  \bibfield  {author} {\bibinfo {author} {\bibfnamefont {J.~F.~L.}\
  \bibnamefont {Duval}},\ }\bibfield  {title} {\enquote {\bibinfo {title}
  {Electrokinetics of diffuse soft interfaces. 2. {A}nalysis based on the
  nonlinear {P}oisson-{B}oltzmann equation},}\ }\href@noop {} {\bibfield
  {journal} {\bibinfo  {journal} {Langmuir}\ }\textbf {\bibinfo {volume}
  {21}},\ \bibinfo {pages} {3247--3258} (\bibinfo {year} {2005})}\BibitemShut
  {NoStop}%
\bibitem [{\citenamefont {Chen}\ and\ \citenamefont {Das}(2015)}]{chen.g:2015}%
  \BibitemOpen
  \bibfield  {author} {\bibinfo {author} {\bibfnamefont {G.}~\bibnamefont
  {Chen}}\ and\ \bibinfo {author} {\bibfnamefont {S.}~\bibnamefont {Das}},\
  }\bibfield  {title} {\enquote {\bibinfo {title} {Streaming potential and
  electroviscous effects in soft nanochannels beyond {D}ebye-{H}{\"u}ckel
  linearization},}\ }\href@noop {} {\bibfield  {journal} {\bibinfo  {journal}
  {J. Colloid Interface Sci}\ }\textbf {\bibinfo {volume} {445}},\ \bibinfo
  {pages} {357--363} (\bibinfo {year} {2015})}\BibitemShut {NoStop}%
\bibitem [{\citenamefont {Silkina}\ \emph {et~al.}(2020)\citenamefont
  {Silkina}, \citenamefont {Molotilin}, \citenamefont {Maduar},\ and\
  \citenamefont {Vinogradova}}]{silkina.ef:2020}%
  \BibitemOpen
  \bibfield  {author} {\bibinfo {author} {\bibfnamefont {E.~F.}\ \bibnamefont
  {Silkina}}, \bibinfo {author} {\bibfnamefont {T.~Y.}\ \bibnamefont
  {Molotilin}}, \bibinfo {author} {\bibfnamefont {S.~R.}\ \bibnamefont
  {Maduar}}, \ and\ \bibinfo {author} {\bibfnamefont {O.~I.}\ \bibnamefont
  {Vinogradova}},\ }\bibfield  {title} {\enquote {\bibinfo {title} {Ionic
  equilibria and swelling of soft permeable particles in electrolyte
  solutions},}\ }\href@noop {} {\bibfield  {journal} {\bibinfo  {journal} {Soft
  Matter}\ }\textbf {\bibinfo {volume} {16}},\ \bibinfo {pages} {929--938}
  (\bibinfo {year} {2020})}\BibitemShut {NoStop}%
\bibitem [{\citenamefont {Ohshima}(1995)}]{ohshima.h:1995}%
  \BibitemOpen
  \bibfield  {author} {\bibinfo {author} {\bibfnamefont {H.}~\bibnamefont
  {Ohshima}},\ }\bibfield  {title} {\enquote {\bibinfo {title} {Electrophoresis
  of soft particles},}\ }\href@noop {} {\bibfield  {journal} {\bibinfo
  {journal} {Adv. Colloid Interface Sci.}\ }\textbf {\bibinfo {volume} {62}},\
  \bibinfo {pages} {189--235} (\bibinfo {year} {1995})}\BibitemShut {NoStop}%
\bibitem [{\citenamefont {Duval}\ and\ \citenamefont {van
  Leeuwen}(2004)}]{duval.jfl:2004}%
  \BibitemOpen
  \bibfield  {author} {\bibinfo {author} {\bibfnamefont {J.~F.~L.}\
  \bibnamefont {Duval}}\ and\ \bibinfo {author} {\bibfnamefont {H.~P.}\
  \bibnamefont {van Leeuwen}},\ }\bibfield  {title} {\enquote {\bibinfo {title}
  {Electrokinetics of diffuse soft interfaces. 1. {L}imit of low {D}onnan
  potentials},}\ }\href@noop {} {\bibfield  {journal} {\bibinfo  {journal}
  {Langmuir}\ }\textbf {\bibinfo {volume} {20}},\ \bibinfo {pages}
  {10324--10336} (\bibinfo {year} {2004})}\BibitemShut {NoStop}%
\bibitem [{\citenamefont {Ohshima}(2006)}]{ohshima.h:2006}%
  \BibitemOpen
  \bibfield  {author} {\bibinfo {author} {\bibfnamefont {H.}~\bibnamefont
  {Ohshima}},\ }\href@noop {} {\emph {\bibinfo {title} {Theory of colloid and
  interfacial electric phenomena}}}\ (\bibinfo  {publisher} {Elsevier},\
  \bibinfo {year} {2006})\BibitemShut {NoStop}%
\bibitem [{\citenamefont {Sobolev}\ \emph {et~al.}(2017)\citenamefont
  {Sobolev}, \citenamefont {Filippov}, \citenamefont {Vorob'eva},\ and\
  \citenamefont {Sergeeva}}]{sobolev.vd:2017}%
  \BibitemOpen
  \bibfield  {author} {\bibinfo {author} {\bibfnamefont {V.~D.}\ \bibnamefont
  {Sobolev}}, \bibinfo {author} {\bibfnamefont {A.~N.}\ \bibnamefont
  {Filippov}}, \bibinfo {author} {\bibfnamefont {T.~A.}\ \bibnamefont
  {Vorob'eva}}, \ and\ \bibinfo {author} {\bibfnamefont {I.~P.}\ \bibnamefont
  {Sergeeva}},\ }\bibfield  {title} {\enquote {\bibinfo {title} {Determination
  of the surface potential for hollow-fiber membranes by the
  streaming-potential method},}\ }\href@noop {} {\bibfield  {journal} {\bibinfo
   {journal} {Colloid J.}\ }\textbf {\bibinfo {volume} {79}},\ \bibinfo {pages}
  {677--684} (\bibinfo {year} {2017})}\BibitemShut {NoStop}%
\bibitem [{\citenamefont {Andelman}(1995)}]{andelman.d:1995}%
  \BibitemOpen
  \bibfield  {author} {\bibinfo {author} {\bibfnamefont {D.}~\bibnamefont
  {Andelman}},\ }\bibfield  {title} {\enquote {\bibinfo {title} {Electrostatic
  properties of membranes: the {P}oisson-{B}oltzmann theory},}\ }\href@noop {}
  {\bibfield  {journal} {\bibinfo  {journal} {Handbook of biological physics}\
  }\textbf {\bibinfo {volume} {1}},\ \bibinfo {pages} {603--642} (\bibinfo
  {year} {1995})}\BibitemShut {NoStop}%
\bibitem [{\citenamefont {Silkina}\ \emph {et~al.}(2019)\citenamefont
  {Silkina}, \citenamefont {Asmolov},\ and\ \citenamefont
  {Vinogradova}}]{silkina.ef:2019}%
  \BibitemOpen
  \bibfield  {author} {\bibinfo {author} {\bibfnamefont {E.~F.}\ \bibnamefont
  {Silkina}}, \bibinfo {author} {\bibfnamefont {E.~S.}\ \bibnamefont
  {Asmolov}}, \ and\ \bibinfo {author} {\bibfnamefont {O.~I.}\ \bibnamefont
  {Vinogradova}},\ }\bibfield  {title} {\enquote {\bibinfo {title}
  {Electro-osmotic flow in hydrophobic nanochannels},}\ }\href {\doibase
  10.1039/C9CP04259H} {\bibfield  {journal} {\bibinfo  {journal} {Phys. Chem.
  Chem. Phys.}\ }\textbf {\bibinfo {volume} {21}},\ \bibinfo {pages}
  {23036--23043} (\bibinfo {year} {2019})}\BibitemShut {NoStop}%
\bibitem [{\citenamefont {Bader}\ and\ \citenamefont
  {Ascher}(1987)}]{bader.g:1987}%
  \BibitemOpen
  \bibfield  {author} {\bibinfo {author} {\bibfnamefont {G.}~\bibnamefont
  {Bader}}\ and\ \bibinfo {author} {\bibfnamefont {U.}~\bibnamefont {Ascher}},\
  }\bibfield  {title} {\enquote {\bibinfo {title} {A new basis implementation
  for a mixed order boundary value {ODE} solver},}\ }\href@noop {} {\bibfield
  {journal} {\bibinfo  {journal} {SIAM J. Sci. and Stat. Comput.}\ }\textbf
  {\bibinfo {volume} {8}},\ \bibinfo {pages} {483--500} (\bibinfo {year}
  {1987})}\BibitemShut {NoStop}%
\bibitem [{\citenamefont {Maduar}\ \emph {et~al.}(2015)\citenamefont {Maduar},
  \citenamefont {Belyaev}, \citenamefont {Lobaskin},\ and\ \citenamefont
  {Vinogradova}}]{maduar.sr:2015}%
  \BibitemOpen
  \bibfield  {author} {\bibinfo {author} {\bibfnamefont {S.~R.}\ \bibnamefont
  {Maduar}}, \bibinfo {author} {\bibfnamefont {A.~V.}\ \bibnamefont {Belyaev}},
  \bibinfo {author} {\bibfnamefont {V.}~\bibnamefont {Lobaskin}}, \ and\
  \bibinfo {author} {\bibfnamefont {O.~I.}\ \bibnamefont {Vinogradova}},\
  }\bibfield  {title} {\enquote {\bibinfo {title} {Electrohydrodynamics near
  hydrophobic surfaces},}\ }\href@noop {} {\bibfield  {journal} {\bibinfo
  {journal} {Phys. Rev. Lett.}\ }\textbf {\bibinfo {volume} {114}},\ \bibinfo
  {pages} {118301} (\bibinfo {year} {2015})}\BibitemShut {NoStop}%
\bibitem [{\citenamefont {Joly}\ \emph {et~al.}(2004)\citenamefont {Joly},
  \citenamefont {Ybert}, \citenamefont {Trizac},\ and\ \citenamefont
  {Bocquet}}]{joly2004}%
  \BibitemOpen
  \bibfield  {author} {\bibinfo {author} {\bibfnamefont {L.}~\bibnamefont
  {Joly}}, \bibinfo {author} {\bibfnamefont {C.}~\bibnamefont {Ybert}},
  \bibinfo {author} {\bibfnamefont {E.}~\bibnamefont {Trizac}}, \ and\ \bibinfo
  {author} {\bibfnamefont {L.}~\bibnamefont {Bocquet}},\ }\bibfield  {title}
  {\enquote {\bibinfo {title} {Hydrodynamics within the electric double layer
  on slipping surfaces},}\ }\href@noop {} {\bibfield  {journal} {\bibinfo
  {journal} {Phys. Rev. Lett.}\ }\textbf {\bibinfo {volume} {93}},\ \bibinfo
  {pages} {257805} (\bibinfo {year} {2004})}\BibitemShut {NoStop}%
\bibitem [{\citenamefont {Belyaev}\ and\ \citenamefont
  {Vinogradova}(2011)}]{belyaev.av:2011a}%
  \BibitemOpen
  \bibfield  {author} {\bibinfo {author} {\bibfnamefont {A.~V.}\ \bibnamefont
  {Belyaev}}\ and\ \bibinfo {author} {\bibfnamefont {O.~I.}\ \bibnamefont
  {Vinogradova}},\ }\bibfield  {title} {\enquote {\bibinfo {title}
  {Electro-osmosis on anisotropic super-hydrophobic surfaces},}\ }\href@noop {}
  {\bibfield  {journal} {\bibinfo  {journal} {Phys. Rev. Lett.}\ }\textbf
  {\bibinfo {volume} {107}},\ \bibinfo {pages} {098301} (\bibinfo {year}
  {2011})}\BibitemShut {NoStop}%
\bibitem [{\citenamefont {Duval}\ \emph {et~al.}(2009)\citenamefont {Duval},
  \citenamefont {Zimmermann}, \citenamefont {Cordeiro}, \citenamefont {Rein},\
  and\ \citenamefont {Werner}}]{duval.j:2009}%
  \BibitemOpen
  \bibfield  {author} {\bibinfo {author} {\bibfnamefont {J.~F.~L.}\
  \bibnamefont {Duval}}, \bibinfo {author} {\bibfnamefont {R}~\bibnamefont
  {Zimmermann}}, \bibinfo {author} {\bibfnamefont {A~L}\ \bibnamefont
  {Cordeiro}}, \bibinfo {author} {\bibfnamefont {N}~\bibnamefont {Rein}}, \
  and\ \bibinfo {author} {\bibfnamefont {C}~\bibnamefont {Werner}},\ }\bibfield
   {title} {\enquote {\bibinfo {title} {Electrokinetics of diffuse soft
  interfaces. {IV}. {A}nalysis of streaming current measurements at
  thermoresponsive thin films},}\ }\href@noop {} {\bibfield  {journal}
  {\bibinfo  {journal} {Langmuir}\ }\textbf {\bibinfo {volume} {25}},\ \bibinfo
  {pages} {10691--10703} (\bibinfo {year} {2009})}\BibitemShut {NoStop}%
\bibitem [{\citenamefont {Garg}\ \emph {et~al.}(2016)\citenamefont {Garg},
  \citenamefont {Cartier}, \citenamefont {Bishop},\ and\ \citenamefont
  {Velegol}}]{garg.a:2016}%
  \BibitemOpen
  \bibfield  {author} {\bibinfo {author} {\bibfnamefont {A.}~\bibnamefont
  {Garg}}, \bibinfo {author} {\bibfnamefont {C.~A.}\ \bibnamefont {Cartier}},
  \bibinfo {author} {\bibfnamefont {K.~J.~M.}\ \bibnamefont {Bishop}}, \ and\
  \bibinfo {author} {\bibfnamefont {D.}~\bibnamefont {Velegol}},\ }\bibfield
  {title} {\enquote {\bibinfo {title} {Particle zeta potentials remain finite
  in saturated salt solutions},}\ }\href@noop {} {\bibfield  {journal}
  {\bibinfo  {journal} {Langmuir}\ }\textbf {\bibinfo {volume} {32}},\ \bibinfo
  {pages} {11837--11844} (\bibinfo {year} {2016})}\BibitemShut {NoStop}%
\end{thebibliography}%

\end{document}